\documentclass[12pt]{article}

\usepackage{amssymb}

\usepackage[arrow,curve]{xy}

\newtheorem{theorem}{Theorem}

\let\le\leqslant
\let\ge\geqslant

\begin{document}
\title{Solution of the Linear Ordering Problem (NP=P)}
\author{G. Bolotashvili\\
\small{Email: \texttt{bolotashvili@yahoo.com}}}

\date{}

\maketitle

We consider the following problem
\begin{eqnarray}
\max && \sum\limits_{i=1, i \ne j}^n \sum\limits_{j=1}^{n} c_{ij}x_{ij}
\nonumber \\
\textrm{s.~t.}&& 0 \le x_{ij}\le 1, \nonumber\\
&& x_{ij}+x_{ji}=1, \nonumber \\
&& 0\le x_{ij}+x_{jk}-x_{ik} \le 1, i \ne j, \ i \ne k, \ j \ne k,
\ i, j, k =1, ..., n.
 \nonumber
\end{eqnarray}
We denote the corresponding  polytope by $B_n$.
The polytope $B_n$ has integer vertices
corresponding to feasible
solutions of the linear ordering
problem as well as  non-integer vertices.
We denote the polytope of integer vertices as $P_n$.

\

Let us give an example of non-integer vertex in $B_n$ and
describe an exact facet cut. In what follows we will interested
only in generating exact facet cuts.

\

Fig. 1 shows a graph interpretation of a non-integer vertex [1],
$$
\begin{array}{c}
\xy 0;<2mm,0mm>:
(0,12)*{i_1},(0,-2)*{j_1},
(10,12)*{i_2},(10,-2)*{j_2},
(20,10)*{\cdots},(20,0)*{\cdots},
(30,12)*{i_m},(30,-2)*{j_m},
(0,0)*{\bullet}\ar(10,10)*{\bullet}="i2"\ar(30,10)*{\bullet}="im"
\POS(10,0)*{\bullet}\ar(0,10)*{\bullet}="i1"\ar"im"
\POS(30,0)*{\bullet}\ar"i1"\ar"i2"
\endxy\\
\\
\textrm{Fig. 1}
\end{array}
$$
where $\{i_1, ..., i_m\}$, $\{j_1, ..., j_m\}$ $\subset \{1, ...,
n\}$; $\{i_1, ..., i_m\} \cap \{j_1, ..., j_m\}=\emptyset$;
$x_{i_lj_q}=0$, $x_{j_qi_l}=1$, $l\ne q$, $l, q = 1, ..., m;$
the other variables that are not shown at the Figure 1 are equal
to $\frac{1}{2}.$ This is the simplest non-integer vertex of the
polytope $B_n$. For this vertex  all adjacent integer vertices can
be written as:
$$
\arraycolsep.2em
\begin{array}{lll}
\alpha_ki_kj_k\beta_k, &\textrm{where}&\alpha_k \textrm{ is any
ordering from the set }\{j_1, ..., j_m\} \backslash \{j_k\},\\
  &&\beta_k \textrm{ is any ordering from the set }\{i_1, ..., i_m\} \backslash \{i_k\}, \\
\\
\alpha_{kp}i_kj_ki_pj_p\beta_{kp}, &\textrm{where}&\alpha_{kp}
\textrm{ is any ordering from the set }
\{j_1, ..., j_m\} \backslash \{j_k, j_p\}, \\
  &&\beta_{kp} \textrm{ is any ordering from the set }
\{i_1, ..., i_m\} \backslash \{i_k, i_p\},
\end{array}
$$
$$
 k \ne p,\ k, p = 1, ...,m.
$$
All adjacent integer vertices, which number is equal to
\begin{eqnarray}
m \left[(m-1)!\right]^2 +
\frac{m(m-1)}{2}\left[(m-2)!\right]^2
\nonumber
\end{eqnarray}
lie in one hyperplane
\begin{eqnarray}
f(x)=2 \sum\limits_{l=1}^{m} x_{i_lj_l} -
\sum_{l=1}^m \sum_{q=1}^mx_{i_lj_q}= 1.
\nonumber
\end{eqnarray}
This hyperplane for $f(x)\leq 1$ is a facet of the polytope $P_n.$

\

Our aim is to determine exact facet cuts for any non-integer vertex of $B_n$
(and not only for them) in an analogous fashion.

\

Figures 2 and 3 show non-integer vertices
of the polytope $B_n$:
$$
\arraycolsep25pt
\begin{array}{cc}
\xy 0;<12mm,0mm>:<0mm,7mm>::
(0,3.5)*{1},
(2,3.5)*{2},
(4,3.5)*{3},
(0,-.5)*{4},
(2,-.5)*{5},
(4,-.5)*{6},
(1,-.5)*{7},
(3,-.5)*{8},
(0,0)*{\bullet}\ar(2,3)*{\bullet}="2"\ar(4,3)*{\bullet}="3"
\POS(2,0)*{\bullet}="5"\ar(0,3)*{\bullet}="1"\ar"3"\ar(3,0)*{\bullet}="8"
\POS(4,0)*{\bullet}\ar"1"\ar"2"
\POS(1,0)*{\bullet}\ar"1"\ar"3"\ar"5"\ar@/_.7ex/"8"
\POS"8"\ar"1"\ar"3"
\endxy
&
\xy 0;<12mm,0mm>:<0mm,7mm>::
(4,.5)*{1},
(0,.5)*{2},
(2,1.5)*{3},
(1,3.5)*{4},
(3,3.5)*{5},
(2,-.5)*{6},
(4,1)*{\bullet}="1"\ar@/_1ex/(1,3)*{\bullet}="4"
\POS(0,1)*{\bullet}="2"\ar@/^1ex/(3,3)*{\bullet}="5"
\POS(2,2)*{\bullet}\ar"4"\ar"5"
\POS(2,0)*{\bullet}\ar"1"\ar"2"\ar"4"\ar"5"
\endxy\\
\\
\textrm{Fig. 2}&\textrm{Fig. 3}
\end{array}
$$
Noninteger vertex at Figure 2 has an oriented chain
 $7583$ of the length 3,
and non-integer vertex at Figure 3 has an oriented
$614$ chain of the length 2. The oriented chain
$758$ at Figure 2 is independent, that is if we exchange the chain
$758$ with any other chain the rest of the graph does not
change,
while the chains $81$ at Figure 2 and $614$ at
Figure 3 are dependent. We define corresponding dependent and independent oriented chains.

\

The following Theorems take place.
\begin{theorem}\label{t1}
Let $x^0$ be a noninteger vertex in $B_n$ and
assume that in graph interpretation there is a graph
vertex $i$ which is the begin or the end of all adjacent arcs.
Assume that the vertex $i$ can be repeated arbitrarily many times
such that each of the new vertices has the same location
with the other part of the graph as the vertex $i$.
Then the new noninteger vertex, corresponding to the new graph, is
a noninteger vertex of $B_n$, and in the new graph the vertices $i$
and new inserted vertices may be put in any order.
\end{theorem}

\begin{theorem}\label{t2}
Let $x^0$ be a noninteger vertex in $B_n$.
Then there does not exist
corresponding  dependent oriented chains of the length 4.
\end{theorem}

The polytope $B_n$ has noninteger vertices whose fractional
components are equal to $\frac{l}{r}$, $r \ge 2,$ $l<r$, as well.

\

For $r=2$ after matrix transformation we get in all cases the
following non unimodular minimal standard matrix:
\begin{eqnarray}
\left(
\begin{array}{rrr}
1 & -1 & 0 \\
1 & 0 & -1\\
0 & 1 & 1\\
\end{array}
\right).
\nonumber
\end{eqnarray}
For $r=3$ after matrix transformation we get in all cases the
known combination of two minimal standard matrices:
\begin{eqnarray}
\left(
\begin{array}{rrrrr}
1 & -1 & 0& -1 & 0 \\
1 & 0 & -1& 0 & 0\\
0 & 1 & 1 & 0 & 0 \\
1 & 0 & 0 & 0 & -1\\
0 & 0 & 0 & 1 & 1\\
\end{array}
\right).
\nonumber
\end{eqnarray}
For any $r$ after matrix transformation we get  known combination
of $r-1$ minimal standard matrices.

\begin{theorem}\label{t3}
Let $x^0$ be a noninteger vertex of $B_n$, which has fractional components
$\frac{l_1}{r_1}$, $l_1<r_1$, $r_1 \ge 3$. Then we pass to an adjacent
noninteger vertex with fractional
components $\frac{l_2}{r_2},$ $r_2 < r_1$,  $l_2<r_2,$ by
changing an equality
in a basis (thus changing one or more
minimal standard matrices).
\end{theorem}

Let $I_1$, $I_2$, ..., $I_s$ $\subset  \{1, ..., n\}$, and assume
that each set $I_p$, $p=1, ..., s,$ corresponds to noninteger
components of a vertex. Then for each $I_p$, $p=1, ..., s,$ we can
construct a facet cut. If $s=1$ a noninteger vertex is called
simple. A noninteger vertex is  called complex if $s \ge 2.$

\

Thus we have given a general description
of the polytope $B_n.$
\begin{theorem}\label{t4}
Let $x^0$ be a simple noninteger vertex of the polytope $B_n$ with
fractional components $\frac{1}{2}$, assume further that there
does not exist dependent oriented chains with the length 3. Then
all adjacent integer vertices lie in one hyperplane, this
hyperplane is a facet of the polytope $P_n$, and it can be
constructed by a polynomial algorithm.
\end{theorem}

Now we describe the principle for constructing facets.

\

Consider a noninteger vertex $x^0.$ It can be defined
as the solution of the following
 system of the linear basic equalities
\begin{eqnarray}
&& x_{i_lj_l}=0, \ l=1, ..., q; \label{1}\\
&& x_{i_lj_l} + x_{j_lk_l} - x_{i_l k_l} =0,
\ l =q+1, ..., \frac{n^2-n}{2}.
\nonumber
\end{eqnarray}
We introduce artificial variables
$x_{j_l{n+1}}=0,$ $x_{i_l{n+1}}=0,$ into the first
$q$ equalities of the system (\ref{1}):
\begin{eqnarray}
&& x_{i_lj_l}+x_{j_l{n+1}}-x_{i_l{n+1}} =0, \ l=1, ..., q. \nonumber
\nonumber
\end{eqnarray}
With the help of the notation
\begin{eqnarray}
x_{i_lj_l} + x_{j_lk_l} - x_{i_l k_l}
:= x(i_l, j_l, k_l),  \ l =1, ..., \frac{n^2-n}{2},
\nonumber
\end{eqnarray}
we rewrite the system (\ref{1}) as follows:
\begin{eqnarray}
x(i_l, j_l, k_l)=0,  \ l =1, ..., \frac{n^2-n}{2}.
\nonumber
\end{eqnarray}
We can determine $\frac{n^2-n}{2}$ linear independent
adjacent integer vertices
\begin{eqnarray}
x^q(i_s, j_s, k_s)=\delta^q(i_s, j_s, k_s),  \ s, q =1, ..., \frac{n^2-n}{2},
\nonumber
\end{eqnarray}
where $\delta^q$ is either 1 or 0. We can prove that all adjacent
integer vertices lie in the hyperplane:
\begin{eqnarray}
f(x)=
\left|
\begin{array}{rrrr}
x(i_1, j_1, k_1) & \dots & x(i_m, j_m, k_m) & 1 \\
\delta^1(i_1, j_1, k_1) & \dots & \delta^1(i_m, j_m, k_m) & 1 \\
& \dots & & \\
\delta^m(i_1, j_1, k_1) & \dots & \delta^m(i_m, j_m, k_m) & 1 \\
\end{array}
\right|=0 \nonumber
\end{eqnarray}
where $m=\frac{n^2-n}{2}.$
\begin{theorem}\label{t5}
Let $x^0$ be a simple noninteger vertex of the polytope $B_n$ with
fractional components $\frac{1}{2}$, assume further that there
exist $\tau$ dependent oriented chains with the length 3. Then all
adjacent integer vertices lie in $2^\tau$ hyperplanes, each of
them is a facet of the polytope $P_n$, and  can be constructed by
a polynomial algorithm.
\end{theorem}
\begin{theorem}\label{t6}
Let $x^0$ be a simple noninteger vertex of the polytope $B_n$
with fractional components $\frac{l}{r},$ $r \ge3$, $l<r.$ Then
we can construct  all minimal standard matrices and corresponding
noninteger vertices with
fractional components $\frac{1}{2}.$ For every such vertex
we can construct facet cuts.
\end{theorem}
\begin{theorem}\label{t7}
Let $x^0$ be a complex noninteger vertex of the polytope $B_n,$
and $I_1$, $I_2,$ ..., $I_s$ correspond to noninteger values. Then
we can construct facet cuts for each set $I_p$, $p=1, ...,s.$
\end{theorem}
Assume we have generated facet cuts. Solving the problem again we
get the noninteger vertex $x^1$ of the polytope
\begin{eqnarray}
&& 0 \le x_{ij}\le 1, \nonumber\\
&& x_{ij}+x_{ji}=1, \nonumber \\
&& 0\le x_{ij}+x_{jk}-x_{ik} \le 1, i \ne j, \ i \ne k, \ j \ne k,
\ i,
 j, k =1, ..., n, \nonumber \\
 \nonumber
 && f_s^1 \le f_s(x) \le f_s^2, \ s=1, ..., q.
 \end{eqnarray}
Without loss of generality we may assume that
the noninteger vertex $x^1$ satisfies the following linear
system:
\begin{eqnarray}
&& x(i_s, j_s, k_s) =1, \ i=1, ..., p, \nonumber \\
&& f_s(x)=f_s^2, \ s=1, ..., q.
\nonumber
\end{eqnarray}
Now we find all adjacent integer vertices. If all of them lie in
one hyperplane and this hyperplane is a facet of $P_n$ then we
generate this facet and re-solve the problem with a new facet. In
case we cannot determine the facet we solve the auxiliary problem:
$$
\begin{array}{rl}
\max&\left( \sum_{s=1}^{p} x(i_s, j_s, k_s) + \sum_{s=1}^{q}
\frac{f_s(x)-f_s^1}{f_s^2-f_s^1}\right), \\
& 0 \le x_{ij}\le 1,\\
& x_{ij}+x_{ji}=1, \\
& 0\le x_{ij}+x_{jk}-x_{ik} \le 1, i \ne j, \ i \ne k, \ j \ne k,
\ i, j, k =1, ..., n,
\end{array}
$$
With the solution of this problem we can construct the facet of
the polytope $P_n.$ In the case of theorems \ref{t5} and \ref{t6}
we can construct the necessary facets by means of a polynomial
algorithm.

\bibliographystyle{plain}

\end{document}